\documentclass[amssymb, nobibnotes, aps,prl,twocolumn]{revtex4-1}
\usepackage{graphicx,epsfig,grffile}
\usepackage{amsmath} 
\usepackage{bm}
\begin{document}

\title{Catching Conical Intersections in the Act; Monitoring Transient Electronic Coherences by Attosecond Stimulated X-Ray Raman Signals}

\author{Markus Kowalewski}
\thanks{These authors contributed equally to this manuscript}
\author{Kochise Bennett\textsuperscript{*}}
\author{Konstantin E. Dorfman}
\author{Shaul Mukamel}
\email{smukamel@uci.edu}
\affiliation{Department of Chemistry, University of California, Irvine,
California 92697-2025, USA}
\date{\today}%

\begin{abstract}
Conical intersections (CoIn) dominate the pathways and outcomes of virtually all
photophysical and photochemical molecular processes.
Despite extensive experimental and theoretical effort, CoIns have not been directly observed yet and the experimental evidence is being inferred from fast reaction rates and some vibrational signatures.
We show that short X-ray (rather than optical) pulses can directly detect the passage through a CoIn with the adequate  temporal and spectral sensitivity.
The technique is based on a coherent Raman process that employs a
composite femtosecond/attosecond X-ray pulse to detect the electronic coherences 
(rather than populations) that are generated as the system passes through the CoIn.
\end{abstract}

\maketitle

\section{Introduction}
The photochemistry of molecules is of considerable fundamental  interest with direct impact on
synthesis \cite{Hoff12}, chemical sensors \cite{garcia2008optical}, and biological processes \cite{Polli10,Rinaldi14,Barbatti:JPhotochemPhotobiol:2007,Barbatti10,Sobolewski:PCCP:2002}.
Conical intersections (CoIns) of electronic states provide a fast, sub-100-femtosecond
non-radiative pathway that controls product yields and rates in virtually all photochemical and photo-physical processes. At a CoIn, electronic and nuclear degrees of freedom become strongly coupled
and the Born-Oppenheimer approximation, which allowed their separation, breaks down.
Strong experimental evidence for CoIns is based on the observation of fast conversion rates or
other indirect signatures (e.g., suppression of vibrational absorption peaks \cite{Raab98jcp}).
However, their direct experimental observation has not been reported yet. The main obstacle is the rapidly decreasing electronic energy gap during the dynamics, requiring an unusual combination of temporal and spectral resolutions which is not available via conventional femtosecond optical and infrared experiments \cite{Polli10,Horio09,Oliver14,McFarland14}.

We propose a novel, background-free technique that can directly and unambiguously monitor the passage through a CoIn by using recently-developed attosecond broadband X-ray sources.
Available optical techniques monitor state populations \cite{Polli10,McFarland14} or look for signatures in transient vibrational spectra to identify CoIns \cite{Raab98jcp,Oliver14,Timmers14prl,Kowalewski15jcp}.
The technique proposed in this paper looks directly at electronic Raman resonances created by the electronic coherence generated as the system passes through the CoIn and is not sensitive to electronic populations.
The time-dependent energy splitting between the two adiabatic surfaces as well as the phase of the wave function can be directly read off the Raman shift between gain and loss features in the Stokes and anti-Stokes signals.
Simulations demonstrate how this new method allows the precise timing of when and how a nuclear wave packet reaches and passes through the CoIn.

\section{TRUECARS}
\begin{figure*}
\includegraphics[width=\textwidth]{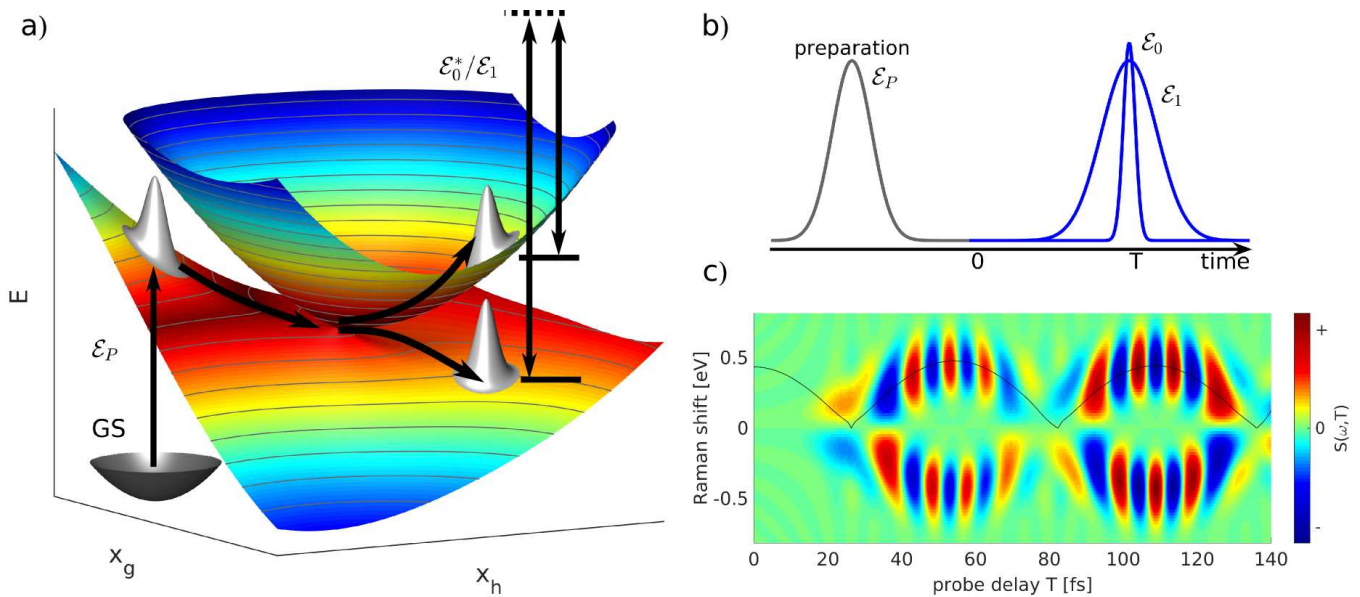}
\caption{Schematic representation of the TRUECARS detection scheme.
(a) A nuclear wave packet is promoted from the ground state (GS) by a pump-pulse $\mathcal{E}_P$ to an excited electronic state. As it passes the coupling region around the CoIn, a coherence is created between the two electronic states.
The broadband $\mathcal{E}_0$/narrowband $\mathcal{E}_1$ hybrid pulse probes the electronic coherence between the nuclear wave packets on different surfaces.
(b) Schematics of the pump and hybrid-probe pulse sequence.
(c) Illustration of the signal calculated for a one-dimensional nuclear model.
The energy splitting of the electronic states involved in the coherence (solid line) can be read from Raman shift.}
\label{fig:diag}
\end{figure*}

Any direct measurement of CoIns simultaneously requires ultra-fast time resolution
and adequate spectral resolution in order to resolve the time dependent electronic energy gap.
As the nuclei approach a CoIn from the vertical transition Franck-Condon point of an optical excitation (Fig. \ref{fig:diag}(a)),
they acquire large velocities and the passage through the CoIn or a seam occurs in
a few femtoseconds \cite{Tao11,Horio09,Patchkovskii14,Kochman15jctc}.
With the ongoing development of free-electron lasers (FELs) \cite{ding2009generation,Helml14nphot} and high-harmonic-generation (HHG) sources \cite{Pop10}, (near transform limited) pulses in the extreme UV to soft X-ray region with a few femtoseconds and down to attosecond durations and several-electron-volt bandwidth \cite{harmand2013achieving, grguravs2012ultrafast, bostedt2013ultra, popmintchev2012bright,manzoni2014coherent} become available.
This makes it possible to directly probe CoIns.

The TRUECARS (Transient Redistribution of Ultrafast Electronic Coherences in Attosecond Raman Signals) technique proposed here is a novel extension of time-domain coherent anti-Stokes Raman spectroscopy (CARS) \cite{lau78,yos94,ideguchi2013coherent, Frostig15}, commonly used to probe vibrational coherence.  In CARS, a pair of optical pulses generates a coherence between vibrational states which is subsequently detected via a Raman process induced by a second pair of pulses.  The detected spectrum is displayed versus the time delay $T$ between the two pairs of pulses, revealing the time-dependent vibrational coherence and its dephasing.  The temporal and spectral resolution may be enhanced by taking the second pulse pair to be a hybrid pulse -- a combination of a narrowband (picosecond) and a broadband (femtosecond) pulse which is known as hybrid CARS \cite{pes08,mil11}.  
\par
The TRUECARS technique, sketched in Fig.~\ref{fig:diag}, extends hybrid CARS in two important respects: 
(i) A combination of attosecond/femtosecond X-ray pulses is used to probe electronic coherence rather than conventional optical pulses that probe vibrational coherence. 
(ii) The coherence is not created directly by applied
pulses as in CARS but is generated internally by the propagation through the CoIn following photoexcitation.
A pump pulse first brings the molecule into an excited electronic state, preparing a non-stationary nuclear wave packet which then propagates towards the CoIn. The electronic coherence is not generated directly by the pulse but instead builds up during the time-evolution of the wave packet as it approaches the vicinity of the CoIn where the non-adiabatic intersurface coupling is present. A hybrid broadband/narrowband X-ray pulse then probes this electronic coherence by the time-resolved gain and loss of the positive and negative stimulated Raman components (see Fig.~\ref{fig:diag}(b) for depictions of the pulse sequence).
Resolving the entire spectrum of electronic Raman transitions (Fig.~\ref{fig:diag}(c)) requires pulses with a few-electronvolt bandwidth and observing the CoIn dynamics requires pulses with a duration on the order of few femtoseconds or less.
Only X-ray pulses provide the necessary temporal and spectral profiles to detect electronic coherences.
\par

The molecule is coupled to the intensity of the off-resonant probing fields via the electronic polarizability operator $\hat\alpha$.  The matter-probe interaction Hamiltonian is
\begin{align}
\hat{H}_{\text{mp}}(t)=\hat{\alpha} |\mathcal{E}_0(t)+\mathcal{E}_1(t)|^2
\end{align}
where $\mathcal{E}_0$ and $\mathcal{E}_1$ are the attosecond (broadband) and femtosecond (narrowband) components respectively of the probing field.
The off-resonant electronic polarizability $\hat\alpha$ is the transition
polarizability describing the Raman transitions between valence states (this is technically frequency-dependent but taken to be flat over the relevant range of frequencies since we are in the off-resonant regime) . We assume
that the dominant transition dipole moments contributing to $\hat\alpha$ are core-to-valence transitions.
We do not include the photo-ionization processes in the simulations. It has been experimentally shown that X-ray Raman signals can successfully compete with the ionization background \cite{miyabe2015transient,Weninger13prl}.
To simplify the analysis, we assume both  components to have the same carrier frequency $\omega_1$.
The TRUECARS signal is defined as the frequency-dispersed photon number change of the attosecond field and is given by
\begin{align}\label{eq:Sfd}
S(\omega,T)=2\Im\int_{-\infty}^{+\infty} \mathrm{d}t\, &e^{i\omega(t-T)}\mathcal{E}_0^*(\omega)\mathcal{E}_1(t-T) \notag\\ 
&\times\langle\psi(t)\vert\hat{\alpha}\vert\psi(t)\rangle
\end{align}
where $T$ is the time-delay between the probe field and the preparation pulse and $\vert \psi(t)\rangle$ is the total (nuclear and electronic) wavefunction.
The probing fields are assumed to be temporally well-separated from the preparation process. The signal carries a phase factor $e^{i(\phi_1-\phi_0)}$, where $\phi_i$ is the phase of the field $\mathcal{E}_i$.  This factor causes the signal to vanish when averaged over random pulse phases; observation of TRUECARS therefore requires control of the relative pulse phases. Note that terms corresponding to electronic populations do not contribute since they carry no dynamical phase and vanish when taking the imaginary part in Eq.~(\ref{eq:Sfd}).  TRUECARS therefore provides a background-free measurement of electronic coherence.  It is also important to note that, due to the frequency-dispersion of the broadband pulse $\mathcal{E}_0(\omega)$, the field-matter interaction time is limited by the femtosecond pulse envelope $\mathcal{E}_1$.  The temporal and spectral resolutions of the technique are not independent but are Fourier-conjugate pairs, both determined by the corresponding temporal and spectral profiles of the femtosecond pulse $\mathcal{E}_1$.   In order to resolve the changing energy gap along the CoIn, $\mathcal{E}_1$ must be shorter than the dynamics while spectrally narrower than any relevant energy splitting.  For example, resolving  a 0.1\,eV energy difference implies at least a 6.5\,fs pulse duration so dynamics faster than this will not be resolved.
\par
The pulse configuration in TRUECARS is identical to transient absorption. The difference is that the probe pulse
is not resonant with any material transitions and is therefore not absorbed. Instead, there is an oscillatory redistribution of intensity between loss (positive Stokes/negative anti-Stokes) and gain (positive anti-Stokes/negative Stokes)
that can affect the frequency-resolved transient intensity.
The signal is linear in the probe intensity $\mathcal{E}_0\mathcal{E}_1$.
Stimulated Raman spectroscopy (SRS)  \cite{yos99,kuk07,sun08,fingerhut2014probing,Bencivenga15nat}  uses the same pulse sequence but detects the quadratic signal $\mathcal{E}_0^2\mathcal{E}_1^2$. TRUECARS is therefore
phase dependent whereas SRS is phase independent.
The quadratic signal would allow greater resolution, since temporal and spectral resolution could then be set by the broadband and narrowband pulses respectively and would not be Fourier limited \cite{kuk07}.  However, the quadratic signal is typically dominated by contributions stemming from electronic populations \cite{Hua15sd} and it is not therefore a background-free measurement of the electronic coherence.
The linear TRUECARS signal is therefore a much cleaner way to measure the passage through a conical intersection.

\par
Qualitative understanding of the TRUECARS signal can be facilitated by a semi-classical picture.
We expand the electronic wave function in the adiabatic basis and
assume that the nuclei follow the classical equations of motion:
\begin{align}
\vert\psi(t)\rangle=\sum_a c_a(t)e^{-i\int_{-\infty}^t \varepsilon_a(\tau)d\tau}\vert a(t)\rangle
\label{eq:AdExp}
\end{align}
where the instantaneous states $\vert a(t)\rangle$ and energies $\varepsilon_a(t)$ vary with time through their dependence on the nuclei while the coefficients $c_a(t)$ vary due to the non-adiabatic coupling between the electronic surfaces near CoIns. The coherence between the surfaces thus propagates with a time-dependent dynamical phase which generates oscillations in $T$ with evolving period and frequency ($\omega_r$). The energy splitting between the electronic states can thus be read not only from $\omega_r$ but also from the oscillation period in $T$ (as can be seen by inserting Eq.~(\ref{eq:AdExp}) into (\ref{eq:Sfd})).
\par
To clearly point out the unique features of the TRUECARS signal, Fig.~\ref{fig:diag}(c) shows a simulation of a single vibrational mode with a long electronic coherence time.
The model is constructed from two electronic states, which are represented by two displaced harmonic
potentials  and a Gaussian diabatic coupling.
This model can represent e.g. a simple diatomic molecule with an avoided crossing.
A full quantum dynamical wave packet calculation is carried out on a numerical grid with a displaced Gaussian wave packet
as initial condition and the TRUECARS spectrum is calculated according to eq. \ref{eq:Sfd}.
In the absence of electronic coherence, the
signal vanishes (this is the case in the beginning of the dynamics,
Fig.~\ref{fig:diag}(c)).
As the wave packet approaches the non-adiabatic coupling region, an
electronic coherence builds up and the signal appears. After
it has passed the intersection, the splitting between the
states increases again. The signal shows an oscillation of gain and loss features in the Stokes and anti-Stokes regime.  The energy splitting (solid line) can be read directly from the Raman shift $\omega_r=\omega-\omega_1$.
The broadening of the signal in $\omega_r$ is caused by the non-vanishing width of the nuclear wave packet, which covers a range of finite width of the potential energy surface.
The signal builds up on both red-  ($\omega_r<0$) and blue- ($\omega_r>0$) sides of the spectrum, appearing as two oscillating peaks. When the red side is positive and the blue side negative,
the energy flows from the pulse to the molecule and 
the process is of Stokes type while opposite conditions yield an anti-Stokes process. The interaction with the molecule thus redistributes the field photons, either shifting the probe pulse toward the red or the blue side of the spectrum, but the total number of photons is conserved \cite{dorfman2015detecting}. This is due to the off-resonant nature of the Raman probe used here (there is no absorption or stimulated emission) and is the origin of the `Redistribution' in TRUECARS. This also leads to the absence of a Rayleigh peak at $\omega_r=0$, which would come from electronic populations,
making the signal background-free (induced only by electronic coherences).
The signal oscillates with time $T$ back and forth between Stokes and anti-Stokes and the oscillation period corresponds to the coherence period (the oscillations speed up and the positions of the peaks in frequency spread apart mirroring the separation of potential energy surfaces).
The oscillation period in $T$ therefore also reveals the separation of adiabatic potential energy surfaces, while the magnitude of the signal envelope reveals the decay of the electronic coherence.

\section{Simulations and Discussion}
\begin{figure}
\centering
\includegraphics[width=.5\textwidth]{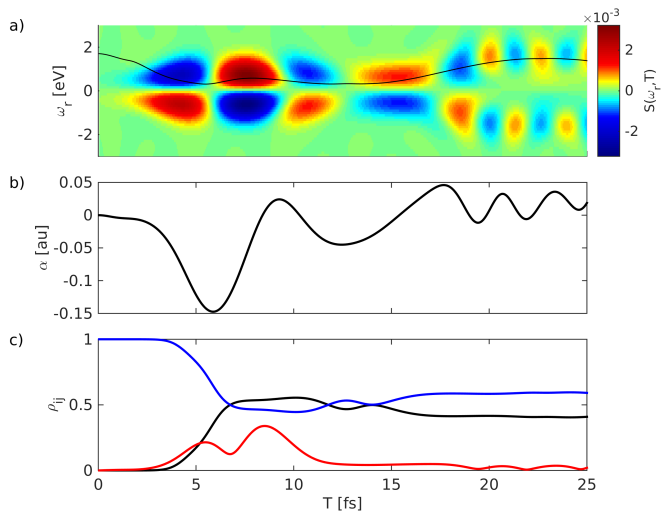}
\caption{(a) Simulated TRUECARS Signal (Eq.~\ref{eq:Sfd}) for the two-dimensional nuclear model
with a pulse length of 1.2\,fs (${\cal E}_1$).
The solid line indicates the average splitting of the potential energy surfaces.
(b) The time dependent expectation value of the polarizability.
(c) Elements of the reduced density matrix of the electronic subsystem.
Blue and black: populations of the adiabatic S$_2$ and S$_1$ state respectively.
Red line: the magnitude of the electronic coherence.}
\label{fig:2DHM}
\end{figure}

We now demonstrate the power of TRUECARS by wave packet simulation on
a more realistic model system with two vibrational modes and two electronic states S$_1$ and $S_2$ and typical molecular parameters (depicted schematically in Fig. \ref{fig:diag}(a)).
This is the minimal model required to describe a CoIn \cite{domcke2004conical}.
The two coordinates resemble the branching space of a CoIn and are displacements
along the derivative coupling vector $x_h$ and the gradient difference vector $x_g$.
The initial condition is at the Franck-Condon point, chosen to be in the vicinity of the CoIn to allow the wave packet to reach the CoIn in a short period of time. Examples of molecules with ultrafast non-adiabatic dynamics include cyclohexadiene \cite{Tamura06jcp}, ethylene  \cite{Tao11}, pyrazine \cite{Horio09jacs},
and DMABN  \cite{Kochman15jctc}.
The wave packet simulations are carried out numerically on a grid in the
electronic and nuclear space using the diabatic basis and are transformed into the adiabatic basis as needed. The details of the calculations are given in the SI.
\par
The molecule is assumed to be initially in its electronic ground state (S$_0$).
An actintic pump-pulse creates an excitation in the
S$_2$ state, thus launching the dynamics.
The diabatic coupling vanishes in the Franck-Condon region to allow for an initial condition in which the Born-Oppenheimer approximation holds.  
The initial  S$_1$/S$_2$  splitting at the Franck-Condon point is around 2\,eV.
The wave packet propagates freely on the S$_2$ surface in the branching space and approaches the CoIn.
The resulting TRUECARS signal (Eq.~\ref{eq:Sfd}) and the averaged time-dependent energy splitting  is shown in Fig.~\ref{fig:2DHM}(a) (solid line).
The qualitative features are similar to the signal from the diatomic model shown in Fig.~\ref{fig:diag}(c). The prepared state contains no electronic coherence
and the signal turns on at around 2\,fs, when the system approaches the non-adiabatic coupling region.
The corresponding molecular property governing the signal, the off-resonant transition polarizability $\alpha(t)$, is shown in Fig.~\ref{fig:2DHM}(b). If the $\hat \alpha$ is assumed to be independent
of the nuclear coordinates, $\alpha(t)$ is directly proportional to the real part of the electronic coherence. In Fig.~\ref{fig:2DHM}(c), the adiabatic populations are shown along
with the magnitude of the electronic coherence.
After the wave packet has passed the CoIn at around 6\,fs,
it travels through a coordinate region where there is a small but finite splitting between adiabatic potential energy surfaces. The signal broadening stems from
two contributions: The width of the nuclear wave packet, covering a certain range
of potential energy differences, and the spectral width of the probe
pulse. The peak maxima are slightly shifted to larger Raman shifts due to the fact
the signal vanishes at $\omega_r=0$ (an effect that is more pronounced for smaller $\omega_r$ as is seen for $T<12$fs in Fig.~\ref{fig:2DHM}(a)). Additionally the information about the energy splitting is
also contained in the oscillations in $T$, indicating that the system is in close vicinity of the CoIn,
as the oscillation frequency is lowered .
At around 15\,fs, the energy splitting increases again as can be seen from $\omega_r$.
Since the S$_1$ and S$_2$ states have different gradients, the overlap between the nuclear wave packets $\langle\Psi_1|\Psi_2\rangle$ decays and the signal fades out.
As can be seen in Fig.~\ref{fig:2DHM}, the passage through the CoIn happens
in less than 12\,fs. By utilizing 1.2\,fs pulses, the
wave packet's arrival at the CoIn can be timed stroboscopically to within 10\,fs.
Note that an even shorter coherence lifetime would not allow for a clear determination
of the energy splitting, but would increase the time resolution.
As is clear from the overlay of the energy splitting on the TRUECARS spectra, the technique is capable of mapping out the potential energy surfaces of the reaction coordinate near the CoIn. It thus gives both dynamical information on the temporal and spectral profile of the the CoIn by providing information about period of oscillations as well as the phase  of the electronic coherences near the CoIn.
TRUECARS might also be useful to measure the Berry phase \cite{Xiao10rmp}, which so far has elluded detection in chemical systems.

In summary, we have presented a new spectroscopic technique (TRUECARS) that can directly monitor passage through conical intersections. The technique measures the frequency-resolved stimulated Raman scattering of a probe pulse as a function of the time delay $T$ with respect to the pump pulse.
In contrast to existing methods, TRUECARS is only sensitive to electronic coherences  and populations do not contribute, making it uniquely suited to probing passage through CoIns by capturing the electronic coherences generated by non-adiabatic couplings in the CoIn vicinity. We simulated the signal for 1D and 2D vibrational model systems and demonstrated that TRUECARS with attosecond pulses can be used to measure the time-varying energy gap between two electronic states.
The rapidly decreasing energy gap around the CoIn is fully visible in the time resolved spectrum
The decay of the electronic coherences contains information about the difference of the gradients between the electronic states, giving a hint about the geometry of the CoIn. 
To precisely time the CoIn and map the energy differences, a molecular system has to pass a CoIn
which is in the close vicinity to the Franck-Condon point. This makes TRUECARS an ideal tool to investigate ultra-fast, photophysical system dynamics.
The experimental parameters required -- broadband sub-femtosecond pulses of $\sim 100$\,eV or more and spectral widths of several eV -- could be realized in the near future from state of the art free-electron laser sources \cite{Helml14nphot,Huang14prac}.

\begin{acknowledgments}
The support of the Chemical Sciences, Geosciences, and Biosciences division, Office of Basic Energy Sciences, Office of Science, U.S. Department of Energy  as well as from the National Science Foundation (grant CHE-1361516) is gratefully acknowledged.  Kochise Bennett was supported by DOE.
M.K gratefully acknowledges support from the Alexander-von-Humboldt foundation
through the Feodor-Lynen program. We would like to thank the
green planet cluster (NSF Grant CHE-0840513) for allocation of compute resources.
\end{acknowledgments}

\end{document}